\documentclass[12pt,preprint]{aastex}
\usepackage{emulateapj5}

\lefthead{Dewangan \& Griffiths}
\slugcomment{Submitted to ApJ}

\newcommand\asca{{\it ASCA}}

\newcommand\chandra{{\it Chandra}}
\newcommand\rosat{{\it ROSAT}}

\newcommand\xmm{{\it XMM-Newton}}

\newcommand\s{{\rm~s}}
\newcommand\kev{{\rm~keV}}
\newcommand\ev{{\rm~eV}}
\newcommand\etal{et al.}
\newcommand\kms{\ifmmode {\rm~km\ s}^{-1} \else ~km s$^{-1}$\fi}
\newcommand\Hunit{\ifmmode {\rm~km\ s}^{-1}\ {\rm Mpc}^{-1}
        \else ~km s$^{-1}$ Mpc$^{-1}$\fi}
\newcommand\ctssec{\ifmmode {\rm~count\ s}^{-1} \else ~count s$^{-1}$\fi}
\newcommand\ergsec{\ifmmode {\rm~erg\ s}^{-1} \else
        ~erg s$^{-1}$\fi}
\newcommand\funit{\ifmmode {\rm~erg\ s}^{-1}\;{\rm cm}^{-2} \else
        ~ergs s$^{-1}$ cm$^{-2}$\fi}
\newcommand\phflux{\ifmmode {\rm~photon\ s}^{-1}\;{\rm cm}^{-2}
        \else   ~photon s$^{-1}$ cm$^{-2}$\fi}
\newcommand\efluxA{\ifmmode {\rm~erg\ s}^{-1}\;{\rm cm}^{-2}\;{\rm
        \AA}^{-1} \else ~erg s$^{-1}$ cm$^{-2}$ \AA$^{-1}$\fi}
\newcommand\efluxHz{\ifmmode {\rm~erg\ s}^{-1}\;{\rm cm}^{-2}\;{\rm
        Hz}^{-1} \else ~erg s$^{-1}$ cm$^{-2}$ Hz$^{-1}$\fi}
\newcommand\cc{\ifmmode {\rm~cm}^{-3} \else cm$^{-3}$\fi}
\newcommand\FWHM{\ifmmode {\rm~FWHM} \else ${\rm~FWHM}$\fi}
\newcommand\Msun{\ifmmode M_{\odot} \else $M_{\odot}$\fi}
\newcommand\Lsun{\ifmmode L_{\odot} \else $L_{\odot}$\fi}
\newcommand\ltsim{\raisebox{-.5ex}{$\;\stackrel{<}{\sim}\;$}}

\newcommand\hbeta{\ifmmode {\rm H}\beta \else H$\beta$\fi}
\newcommand\Kalpha{\ifmmode {\rm K}\alpha \else K$\alpha$\fi}
\newcommand\NH{\ifmmode N_{\rm H} \else N$_{\rm H}$\fi}
\usepackage{graphicx}
\usepackage{here}

\begin{document}

\title{\xmm{} view of the prototype narrow-line Seyfert~1 galaxy I~Zw~1}

\author{G. C. Dewangan \& R. E. Griffiths} 
\affil{Department of Physics, Carnegie Mellon University, 5000 Forbes Avenue, Pittsburgh, PA 15213 USA}

\begin{abstract}
  We present results based on an \xmm{} observation of the prototype 
narrow-line Seyfert~1 galaxy I~Zw~1 performed in June 2002. 
The $0.6-12\kev$ spectrum is well described by  a steep power-law ($\Gamma\sim2.3$) and a weak soft blackbody component ($kT \sim 200\ev$) below $2\kev$. The soft X-ray excess emission is featureless and contributes only $\sim 10\%$ to the total X-ray emission in the $0.6-2\kev$ band. There are tentative evidences for an iron iron K$\alpha$ line and an iron K edge, both from high ionized He-like iron. The $0.6-12\kev$ continuum of I~Zw~1 can be explained in terms of a blackbody ($kT\sim40\ev$), thermal comptonization and its reflection off the surface of an ionized accretion disk. We note that characterizing the soft excess copmonent `above a power law continuum' may not be appropriate due to the low energy curvature of a comptonized spectrum. X-ray emission from I~Zw~1 was highly variabile during the observation. I~Zw~1 showed a large, symmetric X-ray flare during which the X-ray luminosity increased by $\Delta L \sim 6\times 10^{43}{\rm~erg~s^{-1}}$ in an interval of $\sim2800\s$ only. The X-ray spectral variability was unusual during the flare.  In contrast to the general observation of spectral steepening with flux from Seyfert galaxies, there is a clear evidence for hardening of the $0.6-10\kev$ spectrum of I~Zw~1 with increasing flux. This hardening is due to a stronger and slightly flatter power-law component at higher flux without any change in the soft X-ray excess component. The large flare, the accompanying spectral flattening
, and the steady nature of the soft excess component may be due to an additional flatter power-law component during the flare possibly arising as a result of ejection of electron plasma and the SSC process.
\end{abstract}


\section{Introduction}
A significant fraction of type 1 active galactic nuclei (AGN) show `soft X-ray excess 
emission' above a power-law continuum, usually identified as the steepening of the X-ray 
continuum below $\sim 2\kev$. This soft excess emission was first observed by 
{\it HEAO-1} (Pravdo et al. 1981), and {\it EXOSAT} (Arnaud et al. 1985; Singh et al. 1985). 
The number of AGN with soft X-ray excess emission or `ultra soft AGN' increased dramatically with the launch of \rosat{}. Boller, Brandt, \& Fink (1996) showed that AGN with steepest soft X-ray spectra in the \rosat{} band tend to lie at the lower end of the H$\beta$ line width distribution. 
These AGN with FWHM$_{\rm H\beta} \ltsim 2000\kms$ are classified as the narrow-line Seyfert~1 (NLS1) galaxies (Osterbrock \& Pogge 1985), and are distinguished from the bulk of the Seyfert~1 galaxies (`broad-line Seyfert~1s' or BLS1s). NLS1s also have strong optical Fe~\small{II} emission.  Detailed X-ray study of NLS1s was carried out with \asca{} which confirmed the soft X-ray emission and, in addition, showed that the $2-10\kev$ continuum slope too is steeper and anti-correlated with the FWHM of the H$\beta$ line (Brandt, Mathur, \& Elvis 1997; Leighly 1999b; Vaughan \etal 1999b). \asca{} observations also showed that many NLS1s have Fe~K$\alpha$ line  arising from highly ionized iron, and hence the accretion disks of NLS1s must be ionized (Dewangan 2002, and references therein; see Ballantyne, Iwasawa, \& Fabian 2001 for ionized reflection model for NLS1s).   

During the \rosat{} and \asca{} era, NLS1s drew attention of most workers due to the extreme variability of their strong soft X-ray emission.  NLS1s are the most extreme X-ray variable objects among the radio-quiet AGN. NLS1s frequently exhibit rapid (doubling time scales of a few hundred seconds) and/or  large (up to a factor of 100) amplitude X-ray variability (e.g., Boller, Brandt, \& Fink 1996; Dewangan et al. 2001,2002). The excess  variance for NLS1s is typically an order of magnitude higher than that observed for BLS1s. (Leighly 1999a; Turner et al 1999). The strong soft X-ray emission of NLS1s is usually attributed to the intrinsic emission from viscous accretion disks (Pounds et al. 1995). However, the variability of the soft X-ray emission has been found to be faster than the characteristic time scales for an accretion disk (Boller et al. 1993). The fastest time scales for an accretion disk are the thermal
time scale, $t_{th}=4/3\alpha^{-1}\omega^{-1}$, on which thermal instabilities grow, and the sonic time scale, $t_s = r/\omega h$, on which the sound waves cross the disk. In the above, $\alpha$ is the viscosity parameter (Shakura \& Synyaev 1973), $\omega$ is the angular velocity and $h$ is the height scale of the disk.
These time scales are about a factor of two larger than the observed variability time
scale for the NLS1 galaxy IRAS~13224-3809 (Boller et al. 1993). The same is true for many NLS1 galaxies. This apparently argues against the accretion disk origin for the strong soft X-ray emission and the nature of the soft X-ray excess emission remains uncertain.

\chandra{} and \xmm{} grating spectroscopy of NLS1s have revealed that the soft X-ray excess emission is a smooth continuum component rather than a blend of emission/absorption features (Turner et al. 2001; Collinge et al 2001; Gondoin et al. 2002; Page et al. 2003).
It is not clear whether the rapid variability of the smooth soft X-ray emission is due to the variability of the soft excess emission above and over the power-law or it is the underlying power-law component which is varying rapidly. 
Based on a 10~day
{\it ASCA} observation of IRAS~13224-3809, Dewangan et al. (2002) suggested that the
rapid variability in the soft X-ray band is likely to be due to the rapid variability
of the power-law emission and there may not be intrinsically rapid variability in the soft excess
emission
above the power-law. This observation, if confirmed further, will have profound effects on constraining models for X-ray emission of NLS1s. However, this requires time resolved X-ray spectroscopy with short time scales which was not possible with the earlier X-ray observatories. \xmm{} having the largest collecting area and broad bandpass is the most suitable for the above purpose.
 
In this paper, we present results based on an \xmm{} observation of I~Zw~1, including time resolved spectroscopy on time scales of $\sim 5000\s$, comparable to the dynamical time scale for a Keplerian accretion disk $t_{dyn} = 3.2\times 10^{3}(\frac{r}{r_g})^{3/2}(\frac{M}{10^7{\rm~M\odot}})\s$. 
I~Zw~1 is the prototype NLS1 galaxy. Its optical spectrum shows exceptionally narrow Balmer emission lines (FWHM$_{\hbeta}\sim 1240\kms$; Boroson \& Green 1992). I~Zw~1 is well known for its strongest optical and UV Fe~\small{II} emission (Boroson \& Green 1992; Laor et al. 1997). Its high absolute luminosity (M$_V = -23.8$ for H$_0=50{\rm~km~s^{-1}~Mpc^{-1}}$, $q_0=0$) makes it the nearest quasar ($z= 0.060875\pm0.000223$) and  is also called as the `narrow-line quasar' -- a high luminosity version of the NLS1 galaxies. 

\section{Observation and Data reduction}
I~Zw~1 was observed by \xmm{} observatory (Jansen et al. 2001)  on 22 June 2002
using the European Photon Imaging Camera (EPIC),
the reflection grating spectrometer (RGS), and the optical monitor
(OM) instruments.  The EPIC MOS \citep{Turneretal01}, and PN \citep{Struderetal01}
cameras were operated in small, and large window modes,
respectively, to alleviate pileup of photons from the source. The EPIC cameras used the medium optical blocking filter.

The raw events were processed and filtered using the most recent
updated calibration database and analysis software ({\tt SAS v5.4.1})
available in October 2003. Events in the bad pixels and those
adjacent pixels were discarded.  Only events with pattern $0-4$ (single and
double) for the PN and $0-12$ for the MOS were selected.  
Examination of PN light curve extracted from the full field but above $12\kev$
showed  flaring particle background in two observation intervals of about $500\s$.
These intervals were excluded from the rest of the analysis.  This
resulted in `good' exposure time of $18.0{\rm~ks}$ for
the PN and $18.7{\rm~ks}$ for the MOS cameras. 

 Source photons were extracted from the PN and MOS data using a circular region of radius $40\arcsec$ centered at the source position.
Associated background photons were extracted from
source free regions near the source. The total number of net source photons
is $1.43\times 10^5$ detected with the PN and $4.3\times10^{4}$ detected with  
each of the MOS cameras.  

\section{Temporal Analysis}

X-ray light curves of I~Zw~1 were extracted from the PN and MOS data using a circular region of $40\arcsec$ centered at the source position and in the soft ($0.4-2\kev$), hard ($2-10\kev$), and full ($0.4-10\kev$) bands. The soft and hard bands were chosen to separate approximately the two spectral components --  a power-law and a soft excess component generally observed from NLS1 galaxies (see e.g., Leighly 1999b). The time bins with sizes of $200\s$ are at least $50\%$ exposed in all the light curves. 
Background light curves were extracted from source free regions with the same bin sizes, exposure requirements, and in the same energy bands as for the source light curves. The background light curves were subtracted from the respective source light curves after appropriate scaling  to compensate for the different areas of the extraction regions. The three light curves extracted from the PN and MOS data, were combined to improve the signal-to-noise ratio. The observed counts correspond to a mean observed flux of $1.9\times10^{-11}{\rm~erg~cm^{-2}~s^{-1}}$, and luminosity of $1.6\times10^{44}{\rm~erg~s^{-1}}$ in the $0.4-10\kev$ band.

\begin{figure*}
  \centering
\includegraphics[width=12cm]{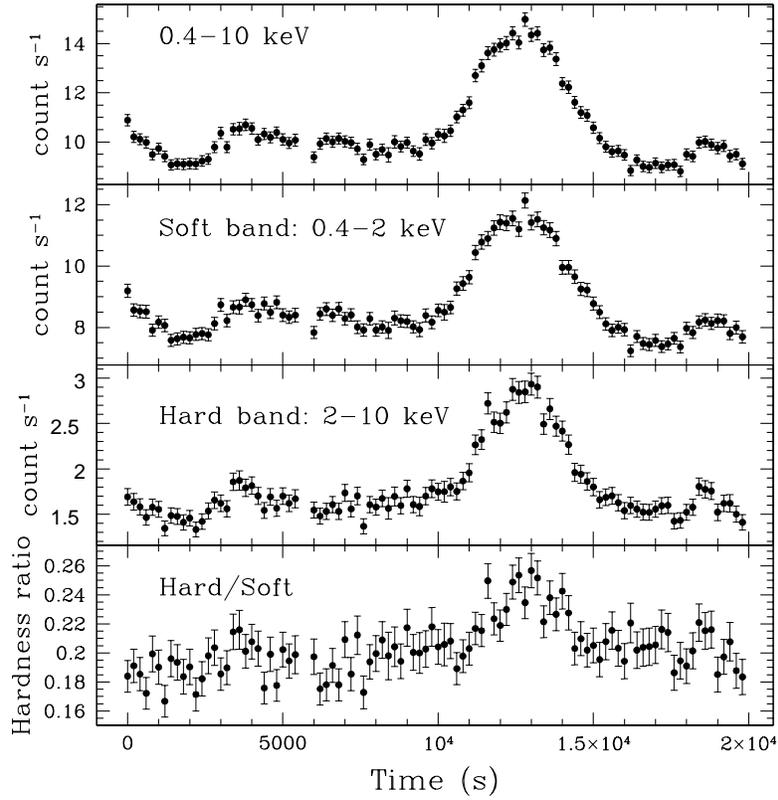}
\caption{Combined PN and MOS light curves of I~Zw~1 in the $0.4-10\kev$, $0.4-2\kev$, and the $2-10\kev$ bands. The bottom panel shows hardness ratio defined as the ratio of count rates in the $2-10\kev$ and
$0.4-2\kev$ bands.}
\label{f1}
\end{figure*}

Figure~\ref{f1} shows the combined PN and MOS light curves of I~Zw~1 in the full, soft, and hard bands. Also shown in Fig~\ref{f1} is the hardness ratio defined as the ratio of count rates in the $2-10\kev$ and $0.4-2\kev$ bands. It is evident that X-ray emission from I~Zw~1 varied strongly during the \xmm{} observation. The light curves show trough-to-peak variations by a factor of $\sim1.6$ in the full band, $\sim1.5$ in the soft band, and $\sim2$ in the hard band. The most remarkable variable event is a large X-ray flare which started at $\sim10^{4}\s$ after the beginning of the observation and lasted for about $5000\s$. During this flare the full band count rate increased from $10.16\pm0.14{\rm~counts~s^{-1}}$ to $14.43\pm0.15{\rm~counts~s^{-1}}$ in just $2800\s$. The count rates quoted above were calculated as the mean of at least three data points.

\begin{figure*}
  \centering
\includegraphics[width=10cm,angle=-90]{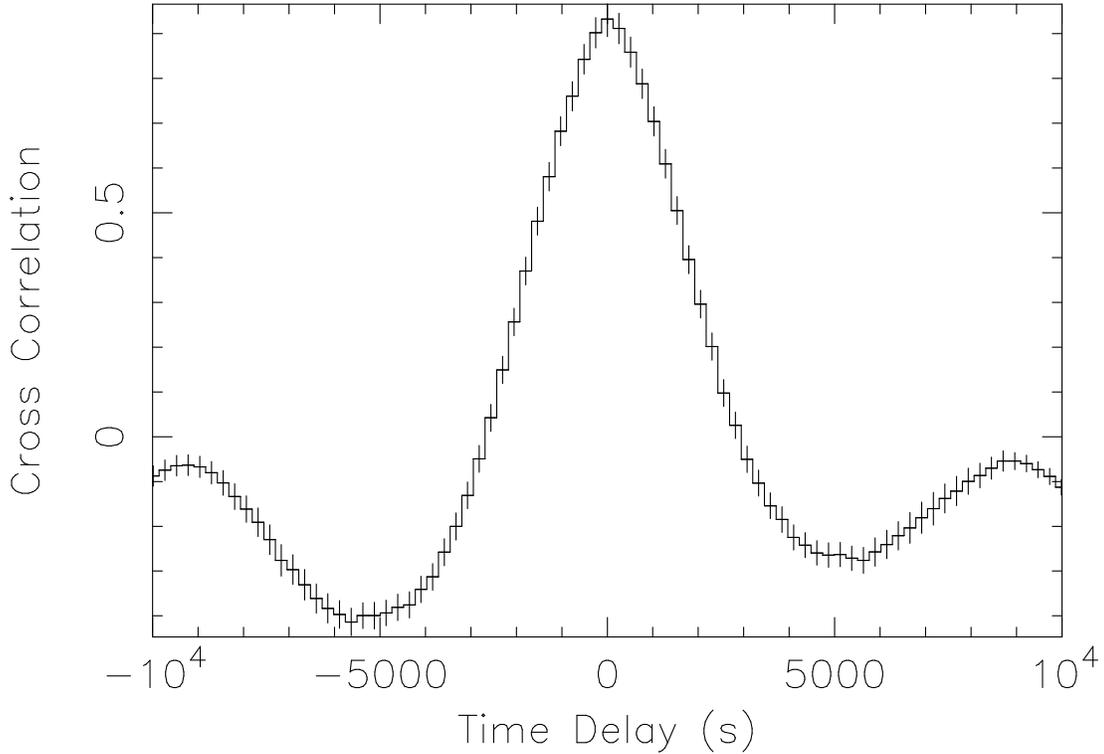}
\caption{Cross-correlation function for the soft ($0.4-2\kev$) and hard ($2-10\kev$) band EPIC light curves of I~Zw~1 with $256\s$ bins. The two bands are strongly
correlated without a delay.}
\label{f3}
\end{figure*}

The soft and hard band light curves show similar variability pattern and appear to be correlated. To quantify this correlation and to search for any possible time lag, we have calculated cross-correlation function (CCF) of the hard band light curve with respect to the soft band light curve. For this purpose we utilized PN and MOS light curves with time bins of $16\s$ created in a similar way as described above. The CCF was calculated using the FTOOLS program {\tt crosscor}. The CCF is plotted in Figure~\ref{f3} with time bins of $256\s$, which shows strong correlation between the soft and hard band light curves without any significant delay. We also examined CCFs with smaller time bins and found no time lag between the two light curves. 

\begin{figure*}
  \centering
\includegraphics[width=10cm]{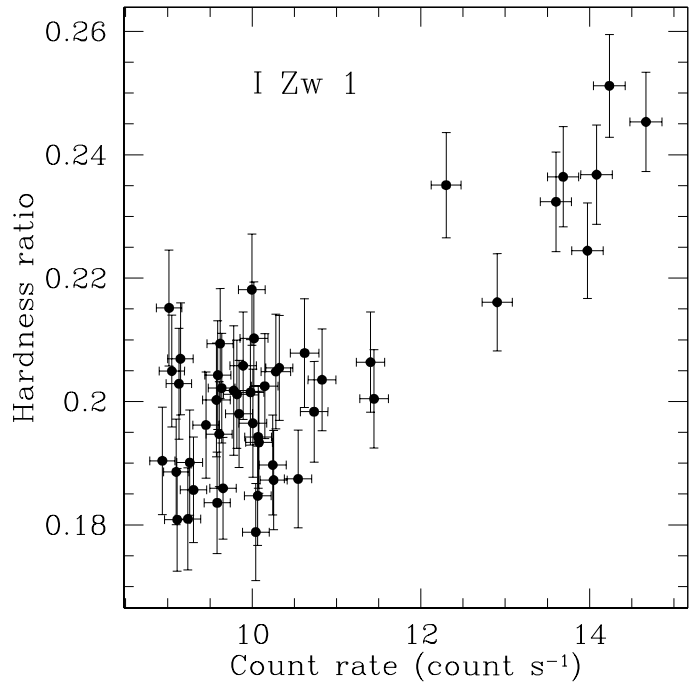}
\caption{Hardness ratio Vs the $0.4-10\kev$ count rate showing spectral hardening
with increasing flux.}
\label{f2}
\end{figure*}

Despite the strong correlation between the soft and hard band light curves, the X-ray flare has different amplitude of variability in the soft and hard bands. The flare is more intense in the hard band (an increase in count rate by a factor of $\sim 1.6$) than in the soft band where the count rate increased by a factor of $\sim1.4$. This is reflected in an increased hardness ratio during the flare (see Fig.~\ref{f1}). It appears that the hardness ratio generally follows the variations in the count rate in any band. This is obvious in Figure~\ref{f2} where the hardness ratio is plotted against the count rate in the full band. Figure~\ref{f2} shows that the X-ray emission from I~Zw~1 hardens with X-ray intensity.   The trend in the hardness ratio with count rate tells us that the total counts in the $2-10\kev$ band increases more than that in the $0.4-2\kev$ and nothing about the $2-10\kev$ or $0.4-2\kev$ spectral change.

\section{Time-averaged spectral analysis}

Photon energy spectra of I~Zw~1 and associated background spectra were accumulated from the EPIC PN and MOS data using the same source and background extraction region described above. The pulse invariant channels were grouped such that each spectral bin contained at least 80 and 30 counts in  the case of PN and MOS spectra, respectively. The EPIC responses were generated using the SAS tasks {\tt rmfgen} and {\tt arfgen}.

The first order RGS1 and RGS2 spectra of I~Zw~1 were extracted using source regions in dispersion/spatial coordinates that contained $90\%$ of the point source events. The corresponding background spectra were generated using the entire regions of CCD arrays but excluding $95\%$ of the point spread function along the spatial direction. RGS response matrices were calculated using the SAS task {\tt rgsrmfgen}. Both the RGS spectra were binned to 5 channels, a further grouping (a minimum of 5 counts per spectral channel) was applied to avoid zero counts in any channel.
  
All the spectral fits  were performed with the XSPEC package (version 11.2; Arnaud 1996) and using the $\chi^2$-statistic for the EPIC spectra and C-statistic for the RGS spectra. Unless otherwise specified, the quoted errors on the best-fit model parameters were calculated for the $90\%$ confidence level for one interesting parameter i.e., $\Delta \chi^2$ or $\Delta C  = 2.71$.  

\subsection{EPIC spectra}
At first the MOS and PN spectra were fitted separately to check for the possible uncertainties due to cross-calibration problems. We found generally good agreement between MOS2 and PN cameras ($\Delta \Gamma \simeq 0.03$) in the $2.5-10\kev$ bands, however, MOS1 results in a flatter ($\Delta\Gamma \simeq 0.15$) spectrum. These discrepancies are consistent with that found by Molendi \& Sembay (2003).  In the $0.6-2.5\kev$ band, MOS1 spectrum is flatter ($\Delta \Gamma \simeq 0.13$) than PN spectrum, while MOS2 data agrees well with the PN data ($\Delta \Gamma \simeq 0.04$).  Therefore we present the spectral results obtained by fitting the same model jointly to the PN and MOS2 data while allowing the relative normalizations to vary. However, PN and MOS data do not agree below $\sim 0.6\kev$. Similar discrepancies at low energies have been found for PG~1211+143 (Pounds et al. 2003) and Akn 564 (Vignali et al. 2003). Therefore, we did not use EPIC data below $0.6\kev$ for out spectral analysis. 

\begin{figure*}
  \centering
\includegraphics[width=5cm,angle=-90]{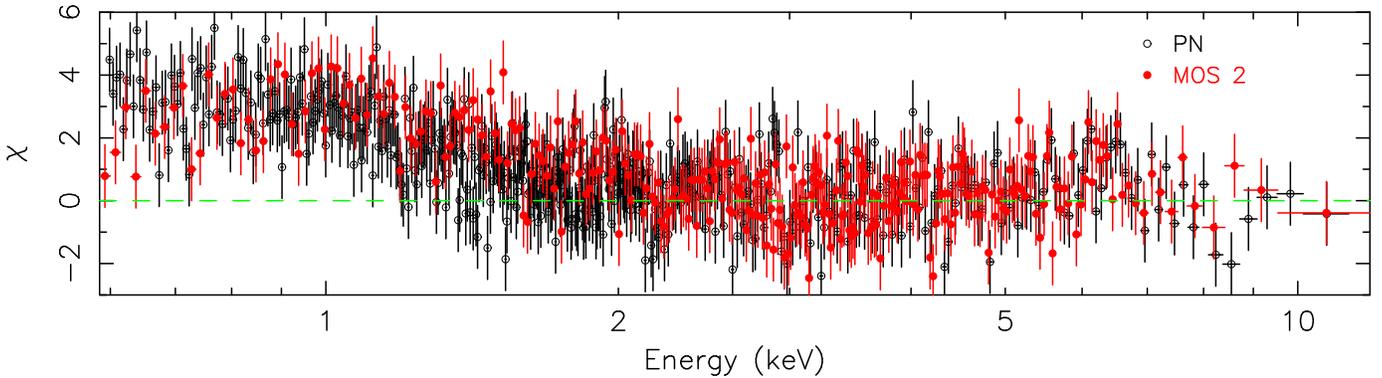}
\caption{A comparison of the observed data in the $0.6-10\kev$ band with the best-fit power-law ($\Gamma \sim 2.28$) model derived using the data in the $2.5-6\kev$ and $7-10\kev$ bands, and extrapolated at lower energies.  The deviations are defined as $\chi = (N_{i}^{\rm obs} - N_{i}^{\rm mod})/\sigma_{i}$, where $N_{i}^{\rm obs}$ is the observed counts in energy channel $i$, $\sigma_i$ is the standard error on  $N_{i}^{\rm obs}$,
and  $N_{i}^{\rm mod}$ is the best-fit
model counts in channel $i$. A weak soft X-ray excess emission below $\sim 2\kev$
and a weak iron $\alpha$ line is seen at $\sim6.7\kev$}
\label{f4}
\end{figure*}

\begin{figure*}
  \centering
\includegraphics[width=5cm,angle=-90]{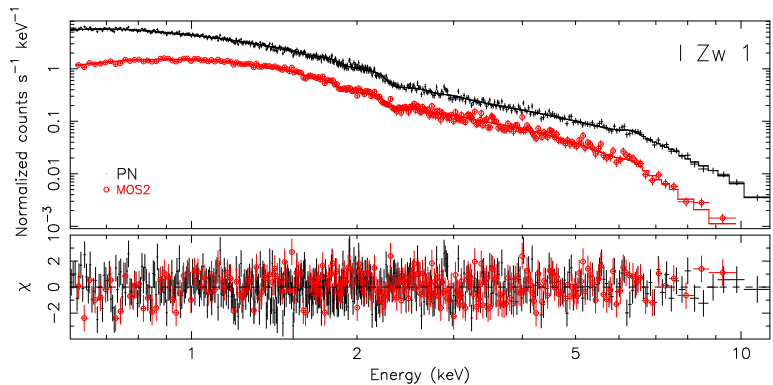}
\caption{PN and MOS2 spectral data and the best-fit model consisting of an absorbed and redshifted blackbody ($kT \sim 235\ev$) and power-law ($\Gamma \sim 2.3$), a Gaussian emission line at $\sim 6.7\kev$, and an absorption edge at
$\sim 8.5\kev$.}
\label{f5}
\end{figure*}

In order to characterize the continuum shape of I~Zw~1, first we fitted a redshifted and absorbed 
power-law model jointly to the PN and MOS2 data. For this fit we chose the energy bands 
$2.5-6\kev$ and $8-10\kev$ bands in the observer's frame. These bands are relatively free of 
any spectral features in the AGN spectra. This model resulted in a good fit ($\chi^2 = 304.7$ for 292 degrees of freedom (dof)). The best-fit power-law photon index is very steep ($\Gamma_X \sim 2.28$). A comparison of the observed data with the extrapolated best-fit power-law to lower energies is shown in Figure~\ref{f4} in terms of deviations  $\chi = (N_{i}^{\rm obs} - N_{i}^{\rm mod})/\sigma_{i}$, where $N_{i}^{\rm obs}$ is the observed counts in energy channel $i$, $\sigma_i$ is the standard error on  $N_{i}^{\rm obs}$, and  $N_{i}^{\rm mod}$ is the best-fit
model counts in channel $i$. Fig.~\ref{f4} shows a weak soft X-ray excess emission above the power-law at energies below $\sim 2\kev$, a weak line-like emission at $\sim 6.7\kev$, and an edge-like feature at $\sim 8\kev$. The soft X-ray excess emission is a general feature of NLS1 galaxies. 
To model this soft component we added a blackbody component to the power-law model and carried out the fitting in the $0.6-10\kev$ band. 
This model provided a good description of the spectra ($\chi^2 = 851$ for $785$ dof). The best-fit absorption column density is $\NH = 8.3_{-0.9}^{+2.2}\times 10^{20}{\rm~cm^{-2}}$, which is significantly higher than the Galactic column density of $N_H^{Gal} = 5.0\times10^{20}{\rm~cm^{-2}}$ (Dickey \& Lockman 1990) along the line of sight to I~Zw~1. This suggests an additional absorber local to the source. Therefore, we fixed the column density to the Galactic value and added an additional photoelectric absorption model at the source redshift to our two component model. This model provided a good fit ($\chi^2 = 846.3$ for $785$ dof) resulted in an excess absorption of $\Delta \NH = 5.1_{-1.6}^{+1.7}\times10^{20}{\rm~cm^{-2}}$.  As also noted earlier, there are signatures of possible presence of an iron K$\alpha$ line and K-edge. Addition of a redshifted  absorption edge to the model consisting of a blackbody and a power-law modified by Galactic and intrinsic absorption improves the fit ($\Delta \chi^2 = -9.3$ for 2 additional parameters). Adding a Gaussian line further improves the fit ($\chi^2 = 822.1$ for $780$ dof). The iron K-edge and the iron K$\alpha$ line are detected at significance levels of $98.6\%$ and $99.7\%$, respectively, according to an F-test. 
 Figure~\ref{f5} shows the observed PN and MOS2 data, the best-fit model, and the deviations $\chi$ (see above) of the data from the best-fit model.
\subsection{RGS spectra}
\begin{figure*}
  \centering
\includegraphics[width=5cm,angle=-90]{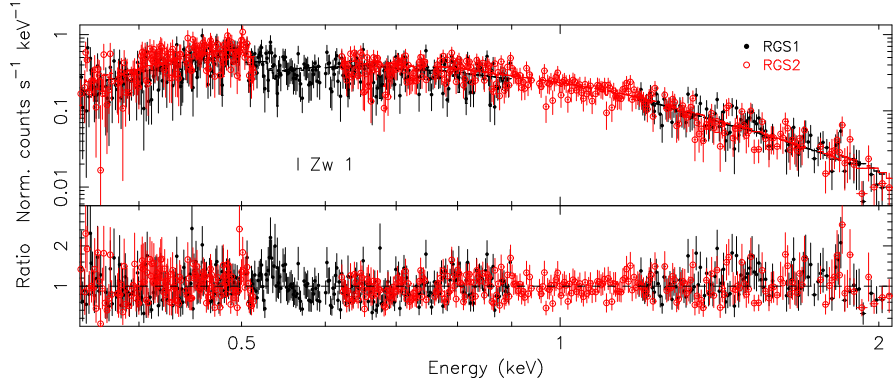}
\caption{RGS spectra of I~Zw~1 and the best-fit absorbed power-law ($\Gamma \sim 2.4$) model in the $0.35-2\kev$ band. The lower panel shows the ratio of the observed data and the best-fit model. The only obvious features are in the $0.5-0.6\kev$ band and near $\sim 1.8\kev$.}
\label{f6}
\end{figure*}

Due to its excellent spectral resolution, the RGS instrument allows a 
more detailed examination of the soft X-ray spectrum. The RGS spectra of 
I~Zw~1 are shown in Figure~\ref{f6}. The soft X-ray spectrum of I~Zw~1 
does not show strong emission or absorption features. It is a smooth 
continuum except for a few weak features. Therefore, we fitted an absorbed power-law model jointly 
to the RGS1 and RGS2 spectra in the $0.35-2\kev$ band. The relative normalizations were allowed to vary. The model resulted in 
C-statistic $= 857.5$ for 805 dof with significant intrinsic (excess over the Galactic column) absorption of $\Delta \NH = 6.3_{-2.0}^{+0.9}\times 10^{20}{\rm~cm^{-2}}$ and a steep power law with $\Gamma = 2.41_{-0.20}^{+0.05}$. The excess absorption is similar to that derived from the PN and MOS data. However, the soft X-ray power-law appears to be slightly steeper ($\Delta \Gamma \sim 0.1$) than the hard X-ray ($2.5-10\kev$) power-law derived from the PN and MOS data. The best-fit model and ratio of the observed data and model are shown in Fig.~\ref{f6}. There are no strong sharp features. The only obvious features are in the $0.5-0.6\kev$ band, probably associated with the detector response of the oxygen K-edge, and emission line-like feature at a rest-frame energy of $\sim 1.9\kev$. Adding a narrow Gaussian line at $\sim 1.94\kev$ results in $\Delta C = -7.3$, and an equivalent width of $\sim 20\ev$.  The rest-frame line centroid energy is not well constrained but appears to be higher than the centroid energy of Si K$\alpha$ line ($1.74\kev$ for neutral Si).

\subsection{The soft X-ray excess emission}
The blackbody and power-law model described the data adequately, but the temperature of the blackbody ($kT \sim 230\ev$) is unrealistically high for a standard accretion disk (see Section 8.4).  Appropriate accretion disk models in place of a simple blackbody result in similar temperatures.
The multicolor disk blackbody  model {\tt diskbb} provided equally good fit ($\chi^2 = 821.3$ for 780 dof) and resulted in $kT= 309_{-49}^{+46}\ev$, while the more realistic disk model {\tt diskpn} (Gierli{\' n}ski et al. 1999) provided equally good fit and resulted in $kT=300_{-49}^{+44}\ev$.

\subsection{Ionized reflection model}
The presence of an iron K$\alpha$ line at $\sim6.7\kev$ and an iron K-edge at $\sim 8.5\kev$ both suggest that these features arise from highly ionized iron (Fe~\small{XXIII} -- Fe~\small{XXV}).
The strength (EW $\sim130\ev$) and width of the iron line (FWHM $\sim 27000\kms$) suggests that the line is 
produced in the accretion disk. These observations are supportive of Compton 
reflection emission from a highly ionized accretion disk. The reflection emission from ionized material can be strong below $\sim 2\kev$  and contribute significantly to the soft X-ray excess emission. With this in mind we adopt an X-ray continuum model that includes a power-law and Compton reflection from ionized accretion disk. We used the Compton reflection model {\it pexriv} of Magdziarz \& Zdziarski (1995) which calculates Compton reflection of illuminated X-ray power-law continuum by an ionized accretion disk. The {\tt pexriv} model does not calculate the iron K$\alpha$ line, therefore we also included a Gaussian line for the iron K$\alpha$ line. As before, we used the photoelectric absorption due to Galactic and intrinsic columns. This model resulted in a good fit ($\chi^2 = 831.1$ for 781 dof). The best-fit parameters were $\Gamma = 2.52_{-0.04}^{+0.05}$, $R = 1.6_{-0.5}^{+0.6}$ (where $R=1$ corresponds to a solid angle of $2\pi$ subtended by the reflector), $\xi = 773_{-773}^{+1478}{\rm~erg~cm~s^{-1}}$. Slightly higher value of $R$ may suggest an additional spectral component for the soft X-ray excess emission. Addition of a blackbody component further improves the fit.
 The model consisting of a blackbody,  power-law, reflection and a  Gaussian, and modified by Galactic as well as intrinsic absorption (model 2)  improves the fit ($\chi^2 = 820.8$ for 779 dof).  The best-fit parameters were $\Gamma = 2.41_{-0.09}^{+0.06}$, $R = 0.65_{-0.47}^{+0.67}$, $\xi = 1683_{-1624}^{+3066}{\rm~erg~cm~s^{-1}}$, and parameters of the Gaussian line were $E = 6.72_{-0.20}^{+0.22}\kev$, $\sigma = 225_{-225}^{+112}\ev $, and $EW = 89$ (see Table~\ref{tab1}). The best-fit model and the unfolded spectrum are shown in Figure~\ref{f6}. Due to the contribution of the ionized reflection in the soft X-ray band, the strength of the blackbody component is weaker in model 2 compared to that derived in model 1. 

\subsection{Ionized reflection and thermal comptonization}
Inverse Compton scattering of soft photons by hot electrons, possibly located in a corona above the disk, can produce approximate power-law continuum. The primary continuum emission of AGN is a comptonized spectrum not an exact power law. The departure at low energies of a thermally comptonized spectrum from a power-law continuum may have profound effects on the shape of the soft X-ray excess emission of NLS1 galaxies. 
In model 2, the soft X-ray excess component is modeled by ionized reflection and a weak blackbody component. Since the thermal comptonized spectrum is an approximate power-law, we further investigate the significance of the weak blackbody component after replacing the power-law component in model 2 by  a comptonization model. We used the thermal comptonization model {\tt comptt} (Titarchuk 1994). The free parameters of the model are the temperature of the seed photons ($kT_{s}$), assumed to follow a Wien law, the optical depth of the scattering region ($\tau$), the electron plasma temperature ($kT_e$), and a parameter defining the geometry (a disk or a sphere). In the following, we consider the disk geometry.  We fixed the parameters of the Gaussian line and the illuminating spectrum in the pexriv model to be the same as that of model 2 but varied the reflection fraction $R$. This model provided a good fit ($\chi^2 = 815.2/783$). The best-fit blackbody temperature is now $38_{-12}^{+45}\ev$. The best-fit relative reflection is $0.8\pm0.1$. The best-fit parameters of the comptt component are seed photon temperature $kT_{s}=155_{-20}^{+5}\ev$, electron temperature $kT_e= 12.2_{-6.3}^{+44.8}\kev$, and $\tau = 1.3_{-0.7}^{+0.2}$.

\subsection{Two component thermal Comptonization model}
A corona with two electron populations with distinct temperatures can give rise to both a soft X-ray excess and a power-law. It is likely that the surface of an accretion disk is ionized, the outermost layers may contain thermal electrons in addition to the hotter electron population above the disk in the corona. Under such conditions, the cool disk photons will be first comptonized by the cooler of the electron population at the surface of the disk, giving rise to the soft excess emission below $\sim 2\kev$. This soft excess emission may be the seed photons for the comptonization by the hotter electrons in the corona (Page et al. 2003).  We investigated if such a process is a viable model for the X-ray spectrum of I~Zw~1.  
First, we tried to fit the power-law component with the {\tt comptt} model. A combination of a blackbody, comptt, a Gaussian line, and an edge model provided a good fit ($\chi^2 = 825.0$ for 779 dof). No intrinsic absorption is required in this model. Thus the power-law component is well described by comptonization of a soft thermal spectrum with temperature $kT=134_{-11}^{+13}\ev$ by hot electrons with $kT_e = 24.4_{-6.7}^{+4.8}$ and the soft X-ray excess component is described by a blackbody component with $kT_{bb} = 292_{-17}^{+21}\ev$. The contribution of this soft excess component to the $0.6-2\kev$ band unabsorbed emission is $\sim 11\%$. As the temperature of the soft excess component is much higher than that expected from an accretion disk. This component is unlikely to be the intrinsic disk emission. Replacing the blackbody compoent by an additional comptonization component the fit slightly ($\chi^2 = 820.7$ for 777 dof). The best-fit parameters are comptt 1: soft photon $kT= 60_{-43}^{+55}$, electron temperature $kT = 16.5_{-4.1}^{+28.2}$ and $\tau =1.7_{-0.2}^{+0.4}$; comptt 2: soft photon $kT= 191_{-19}^{+23}\ev$, electron temperature $kT =29.8_{-18.9}^{+42.2}$ and $\tau =0.29_{-0.16}^{+65}$. In this model, soft thermal photon from the disk with $kT = 60_{-43}^{+55}$ are comptonized by hot thermal ($kT = 16.5_{-4.1}^{+28.2}$) electrons in the corona. This process gives rise to the soft X-ray excess emission. The second comptonization model produces the power-law emission.

\section{Time-resolved spectroscopy}
To examine the spectral variations of I~Zw~1, we extracted 4 time-selected spectra across the 
$20{\rm~ks}$ \xmm{} observation. We chose the sampling time scale of $5000\s$, which separates the flaring interval nicely. Thus we extracted four spectra -- spectrum 1 ($0-5000\s$), spectrum 2($5001-10000\s$), spectrum 3 ($10001-15000$), and spectrum 4 ($15001-20000\s$). 
The resulting ``on-source'' average exposure time was $\sim5~{\rm~ks}$ per EPIC spectrum. We set the ancillary response, and response matrix files to be those of the mean spectrum. Since there is no variation in the background level across the observation, we used the same  background spectra extracted across the full observation for the mean spectra. Again, the energy channels were appropriately grouped to achieve a good signal-to-noise (a minimum of 30 and 20 counts per energy channel for the PN and MOS, respectively). As before, we carried out the spectral fitting
of the time-selected PN and MOS2 data jointly.
                                                                                
\begin{figure*}
  \centering
\includegraphics[width=16cm]{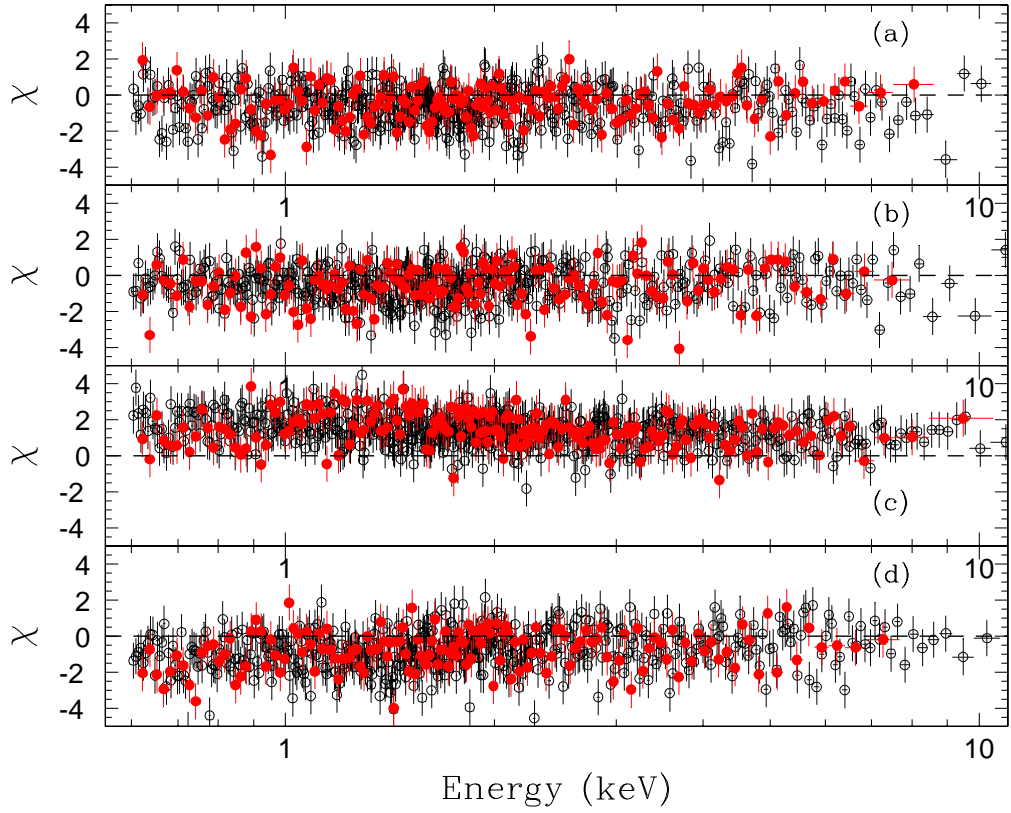}
\caption{Deviations of the individual time-selected PN (open circles) and MOS2 (filled circles) spectra from the best-fit model to the mean spectrum. The time-selected spectra are (a) spectrum 1, (b) spectrum (2), (c) spectrum 3, and (d) spectrum 4.}
\label{f9}
\end{figure*}

\begin{figure*}
  \centering
\includegraphics[width=10cm]{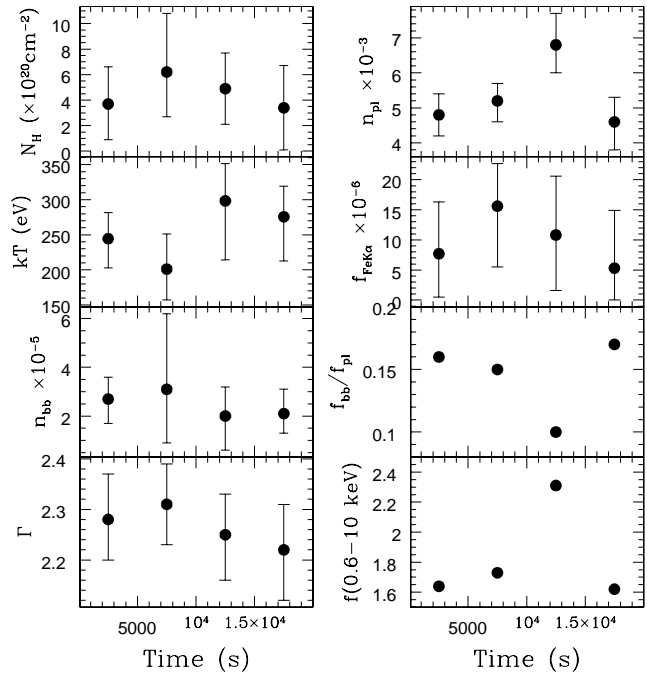}
\caption{Time series of the best-fit spectral parameters derived from the time-resolved spectral fittings. The time bin size is $5000\s$.}
\label{f10}
\end{figure*}

To illustrate the flux and  spectral variability of I~Zw~1, we compared each of
the time-selected spectra with the mean spectrum. Figure~\ref{f9} shows deviations of the time-selected spectra from the best-fit model 1 for the mean spectrum.
These plots were constructed by using the best-fit model 1 for the mean spectrum but without actually fitting the time-selected spectra. Strong variation is present only for the flux. To quantify these variations we performed spectral fitting of the time selected spectra. We used the model 1 (a combination of a blackbody + power law + Gaussian line multiplied by Galactic, intrinsic absorption and
an iron K-edge models) and fitted the 4 time-selected EPIC spectra. It was not possible to constrain both the Gaussian line and iron K-edge for individual spectra. Therefore, we fixed the line energy and width of the Gaussian line and the parameters of the edge model to be the same as the best-fit values derived for the mean spectrum. Model 1 provided good fit to all the the four time-selected spectra of I~Zw~1. The best-fit parameters are listed in Table~\ref{tab2}.
The results of the time-resolved spectroscopy are plotted in Figure~\ref{f10}, which shows that all the parameters except for the power-law normalization changed significantly across the $20{\rm~ks}$ \xmm{} observation. The decrease in the ratio of the flux of the blackbody and power-law components in the $0.6-2\kev$ band during the large flare appears to be due to increase in the power-law flux alone. 

To investigate further, we fit the model 1 to the time-selected PN spectra jointly. We excluded the MOS2 data from this fit to avoid large number of model parameters. As before, the width and
the centroid energy of the iron K$\alpha$ line, and the parameters of the edge model were fixed to the best-fit values derived from the mean spectrum. In this fit, except for the  normalization of the power-law component, all the parameters for a data group were tied to the corresponding parameters in other data groups. The four normalizations of the power-law components for the four time-selected were varied independently. This type of parameterization allowed us to check if the spectral variability of I~Zw~1 is solely due to variations  in the power-law normalization. 
The joint fit to the time-selected spectra resulted in a good fit ($\chi^2 = 1889.1$ for $1903$ dof). Varying the photon indices independently improves the fit ($\Delta \chi^2 = 49.1$ for 3 less dof). There is a marginal evidence for flattening of the power-law during the large flare. The best-fit photon indices are $2.36_{-0.06}^{+0.06}$, $2.33_{-0.06}^{+0.05}$, $2.24_{-0.05}^{+0.04}$, and $2.29_{-0.06}^{+0.05}$ for spectrum 1, 2, 3, and 4, respectively. Varying blackbody temperatures independently does not improve the fit ($\chi^2 = 1834$ for $1897$ dof). The same is true for the blackbody normalization ($\chi^2 = 1835.4$ for 1897 dof).

\section{Flux-selected spectroscopy}
The soft and hard band X-ray emission are correlated without any significant delay. Based on our time-resolved spectroscopy in the preceding section, this correlation appears to be due to the variability of the power-law component alone without any significant changes in the soft X-ray excess component. In order to further explore and to improve the dependence of the spectral parameters, we have carried spectral analysis at low and high flux levels. We extracted two averaged spectra covering the large flare between the elapsed time interval of $\sim 10^4 - 1.5\times10^4\s$ (high flux or flare state) and excluding the large flare (low flux or quiescent state). The mean PN count rates are $6.7\pm0.03{\rm~count~s^{-1}}$ (high flux state) and $5.2\pm0.02{\rm~count~s^{-1}}$ (low flux state) in the $0.6-10\kev$ band. We analyzed both the PN spectra jointly. We used the model 1 (absorbed and redshifted blackbody and  power-law modified by an absorption edge at $\sim 8.5\kev$ and iron K$\alpha$ line at $\sim 6.7\kev$. We fixed the parameters of the iron line and edge to be the same as that derived for the mean spectrum. In order to investigate which model parameter or parameters differ most significantly between the two flux states, we tied all the model parameters for the high flux state to that of low flux state and untie the parameters step-by-step.

\begin{figure*}
  \centering
\includegraphics[angle=-90,width=10cm]{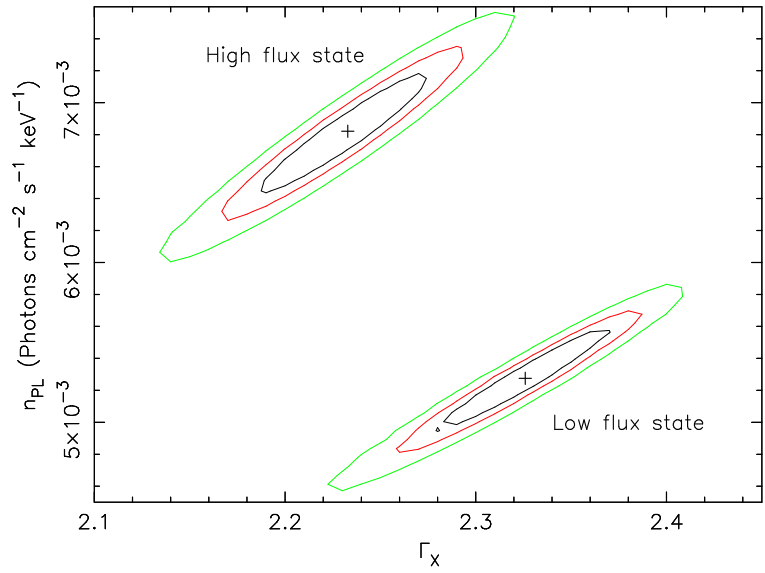}
\caption{Confidence contours of the power-law photon index and normalization for the low and high flux state spectra. The contours are drawn at the confidence levels of $68\%$, $90\%$, and $99\%$, and the best-fit values are marked by a plus sign.}
\label{f11}
\end{figure*}

The best-fit model with the same model parameters for the two flux states
is not a good fit ($\chi^2 = 2782.8$ for $1281$ dof). Varying the power-law normalizations independently resulted in an acceptable fit ($\chi^2 = 1347.8$ for 1280 dof). Untying the photon indices further improved the fit ($\chi^2 = 1302.3$ for $1279$ dof), with significantly different best-fit photon indices of $\Gamma = 2.24_{-0.04}^{+0.04}$ (high flux state) and $2.34_{-0.05}^{+0.04}$ (low flux state). We also varied the blackbody normalization independently, the fit did not improve further ($\chi^2 = 1300.0$ for $1278$ dof). Thus we conclude that the blackbody normalization did not vary between the two flux states and tied the parameter again in the following analysis.  
We calculated the confidence contours for the photon indices and the power-law normalizations in the low and high flux states. 
Figure~\ref{f11} shows these contours at $68\%$, $90\%$, and $99\%$ confidence levels.

The brightening and flattening of the power-law during the X-ray flare suggests that the flare could have resulted due to an additional power-law component. To test such a scenario, first we tied the photon index and normalization for the two flux states and introduced a second power-law component for the high flux state. This model resulted in a good fit ($\chi^2 = 1305.2$ for 1279 dof). The best-fit photon index and the normalization  of the second power-law are $\Gamma = 2.01_{-0.07}^{+0.07}$ and $n_{PL} = 1.55_{-0.09}^{+0.15}\times 10^{-3}{\rm~photons~cm^{-2}~s^{-1}}$, respectively.  The first power-law is steeper and stronger ($\Gamma = 2.34_{-0.04}^{+0.08}$; $n_{PL} = 5.36_{-0.26}^{+0.14}\times 10^{-3}{\rm~photons~cm^{-2}~s^{-1}}$). 

\section{Black hole mass for I~Zw~1}
Before we discuss the results below, we estimate the mass of the black hole for I~Zw~1. 
Calculations of the black hole mass rely on the assumption
that the dynamics of the optical broad-line region gas is dominated by the gravitational potential of the central super-massive black
hole. To calculate the black hole mass for I~Zw~1, we used the
results of Kaspi et al. (2000), based on a
reverberation study of 17 quasars.
Kaspi et al. (2000) have determined an  empirical relationship between the size of
the broad-line region
($R_{BLR}$) and the monochromatic luminosity ($\lambda L_{\lambda}$) as follows.
\begin{equation}
R_{BLR} = (32.9) \left[\frac{\lambda L_{\lambda}(5100 {\rm~ \AA})}{10^{44} {\rm~erg~s^{-1}}}\right]^{0.700} {\rm~lt-days}
\end{equation}
We use the flux taken from a linear interpolation between
photometric data points obtained from the {\em Third Reference Catalog of Bright Galaxies} (RC3;
de Vaucouleurs et al. 1991) at the rest wavelength of $5100{\rm~\AA}$ ($\lambda f_{\lambda} (5100{\rm~AA}) = 6.1\times10^{-11}{\rm~erg~cm^{-2}~s^{-1}}$).
The mass of the black hole is given by
$M_{BH}=rv^{2}G^{-1}$. To determine $v$, the velocity, we correct
$v_{FWHM}$ of the H$\beta$ emission line by a factor of $\sqrt{3}/2$ to account for velocities in three dimensions We used $v_{FWHM} = 1240\kms$ for the H$\beta$ line from Boroson \& Green (1992). The mass is then
\begin{equation}
M = 1.464 \times 10^{5} \left(\frac{R_{BLR}}{\rm~lt-days}\right)\left(\frac{v_{FWHM}}{10^3{\rm km~s^{-1}}}\right)^{2} {\rm~M\odot}
\end{equation}
Using the cosmology $q_{0} =0.5$ and $H_{0}=70$km s$^{-1}$Mpc$^{-1}$,
this gives us a value for the central black
hole mass of $2.4 \times 10^7 M_{\odot}$.

\section{Discussion}
This paper presents an analysis of $\sim 20{\rm~ks}$ \xmm{} observation of the prototype narrow-line Seyfert~1 galaxy I~Zw~1. The time-averaged X-ray spectrum of I~Zw~1 shows four spectral components -- a power-law continuum, a soft X-ray excess component below $\sim 2\kev$, weak intrinsic absorption ($N_H \sim 2\times 10^{20}{\rm~cm^{-2}}$), and spectral features due to iron (the K$\alpha$ line and K-edge). Above $2.5\kev$, the power-law component has a slope of $\Gamma \sim 2.3$ which is steeper than that of most BLS1 galaxies ($\Gamma \sim 1.9$) but is similar to other NLS1 galaxies e.g, Ton~S180 ($\Gamma=2.26_{-0.12}^{+0.05}$; Vaughan et al. 2002), Mrk~335 ($\Gamma=2.29\pm0.02$; Gondoin et al.2002); Akn~564 ($\Gamma \simeq 2.50 - 2.55$; Vignali et al. 2003). The X-ray continuum slightly steepens at lower energies, resembling to the soft X-ray excess emission generally observed from NLS1 galaxies. The strength of this soft X-ray emission is weaker in I~Zw~1 than that in many NLS1s. The contribution of the soft X-ray excess emission to the $0.6-2\kev$ band X-ray emission is only $10\%$ in I~Zw~1. \asca{} observation of I~Zw~1 in 1995 also showed intrinsic absoption ($N_H \sim 3\times10^{20}{\rm~cm^{-2}}$), steep power-law continuum ($\Gamma_X \sim 2.4$), and an iron K$\alpha$ line at $\sim7\kev$ but did not detect any soft excess component (Leighly 1999). I~Zw~1 was fainter by a factor of $\sim 2$ during the \asca{} observation. 

\subsection{Rapid variability and the efficiency limit}
True to the nature of NLS1 galaxies, I~Zw~1 showed rapid and large amplitude X-ray variability during the \xmm{} observation. X-ray emission from I~Zw~1 increased by $40\%$ in $2800\s$ only,
 corresponding to a change in the $0.6-10\kev$ luminosity of $\Delta L = 6\times 10^{43}{\rm~erg~s^{-1}}$. 
This is a remarkable variability for a radio-quiet quasar as the time scale is  a factor of $\sim 6$ shorter than the dynamical time scale for a $10^7{\rm~M\odot}$ Schwarzschild black hole. Such variability is rare among radio-quiet quasars. Other radio-quiet quasars to show such rapid and large amplitude variability are PHL 1092 (Brandt et al. 1999), RX J1334.9+3759 (Dewangan et al. 2002). 

The major consequence of rapid and large amplitude variability is that the source has to be emit very efficiently. It is possible to calculate the efficiency of conversion of rest mass into energy using simple arguments originally due to Fabian (1979). If the X-ray emitting region is an accreting spherical cloud of ionized hydrogen, then there is a limit on the radiative efficiency, $\eta$, given by $\eta > 4.8\times 10^{-43} \Delta L/\Delta t$, where $ \Delta L$ is the change in luminosity in the rest-frame time interval $\Delta t$. First, we convert the change in the combined PN$+$MOS count rate into change in the luminosity by using a constant conversion factor of $(1.5\pm0.1)\times10^{43}{\rm~erg~count^{-1}}$, derived from the time averaged PN$+$MOS count rate and the intrinsic luminosity. The change in the count rate of $3.99\pm0.14{\rm~count~s^{-1}}$ over $2800\pm 565.7\s$ ($2639.3\pm533.2\s$ in the rest frame) corresponds to a rate of change of luminosity, $\Delta L/\Delta t = (2.27\pm0.49)\times 10^{40}{\rm~erg~s^{-2}}$. This gives a lower limit to the efficiency, $\eta > 0.011\pm0.002$.
This efficiency can be compared with the maximum efficiency possible for a Schwarzschild black hole ($\eta \simeq 0.06$; Shapiro \& Teukolsky 1983) and a Kerr black hole ($\eta \simeq 0.3$; Thorne 1974). Thus the rapid flare observed from I~Zw~1 is consistent with that expected for a Schwarzschild black hole.

Our time resolved spectroscopy has revealed that the soft X-ray excess emission, presumably arising from an accretion disk, did not vary appreciably during the giant flare seen in Fig.~\ref{f1}. The rapid and large amplitude variability is consistent with a flaring of the power-law component alone. The primary power-law emission of radio-quiet AGN is thought to be the result of comptonization of soft photons from the disk by hot, relativistic electrons in a region, the geometry of which is not clear at present. The possibilities are -- ($i$) an extended corona above a cold accretion disk (e.g, Haardt \& Maraschi 1991), ($ii$) non-thermal electrons at the base of a jet, and ($iii$) thermal electrons in the innermost regions of an accretion disk. 
In the first case, rapid and large amplitude variability can be produced by magnetic flares arising from the surface of the disk.
In the second case, ejection of a relativistic plasma blob as a jet would produce the giant flare with power-law spectral shape. In the third scenario, it is difficult to produce giant and rapid flares.
It is difficult to distinguish between the first two possibilities in the absence of wide band X-ray spectrum up to several hundred keV. However, a Kerr black hole is not required in any of the two possible scenarios. Magnetic flares can be produced on very short time scales as the reconnection time scale can be very short (Merloni \& Fabian 2001).  

\subsection{Flux correlated spectral variability}
There is a clear evidence for a change in the hardness ratio with total count rate (see Fig.~\ref{f2}). The full band X-ray spectrum of I~Zw~1 hardens with increasing flux.
A detailed time-selected spectral analysis revealed that hardening of the full band spectrum as suggested by the hardness ratio is due to the increase in the power-law normalization and a steady soft excess component. Our flux-selected spectroscopy has revealed that the power-law component became stronger and slightly flatter  in the flaring or high flux state. Thus the change in the hardness ratio is not only due to the variations in the $2-10\kev$ band flux   but also due to hardening of the primary power-law.
The spectral hardening of I~Zw~1 with increasing flux is in sharp contrast with that observed from many Seyfert galaxies, which generally show spectral steepening with increasing flux 
(Turner 1987; Singh et al. 1991; Turner et al. 1990; McHardy, Papadakis \& Uttley 1998; Lee et al. 2000; Chiang et al. 2000; Done et al. 2000; Zdziarski \& Grandi 2001; Petrucci et al. 2001; Shih, Iwasawa \& Fabian 2002; Vaughan and Edelson 2001; Nandra 2001; Romano et al. 2002; Dewangan et al. 2002; Gallo et al. 2003; Lamer et al. 2003).  Georgantopoulos \& Papadakis (2001) found similar behavior in a number of Seyfert~2 galaxies.
To explain such spectral variability, McHardy et al.(1998) and Shih et al. (2002) proposed a two component model, consisting of a soft component with constant spectral shape but variable flux and a hard component with constant flux, so that the soft component dominates the spectrum at high fluxes and overall spectrum becomes steeper. Using a long \xmm{} observation, Fabian \& Vaughan (2003) showed that the spectral variability of MCG-6-30-15, on time scales of $10{\rm~ks}$, is characterized by a soft varying component which is a power-law with constant slope and a hard constant component produced by very strong reflection from a relativistic disk.  In contrast, in the case of I~Zw~1, it is the soft component with constant flux and the hard component with variable flux resulting in spectral hardening.  

\subsection{Possible origin of spectral hardening}
The observed $2-10\kev$ spectral steepening with flux has been attributed to the change in the shape of the comptonized spectrum possible due to Compton cooling by increased seed or soft photon flux from the disk (Haardt, Maraschi, \& Ghisellini 1997) or due to increasing covering fraction of the corona resulting from correlated magnetic flare or reconnection events ('thunder-cloud' model of  Merloni \& Fabian 2001). Obviously both the scenarios are not applicable to the spectral behavior of I~Zw~1. The soft excess emission did not increase with flux during the flare, implying that there was no change in either the seed photons from the disk or reprocessed emission by the disk. The soft excess emission attributed to the intrinsic disk emission and reprocessed emission simply did not appear to respond the large X-ray flare. This is possible if the flare emission was beamed away from the accretion disk.  The X-ray flare and accompanied spectral hardening of I~Zw~1 is suggestive of a generation of a new population of hot or relativistic electrons without an increase in the covering fraction of the corona. This new population must be short-lived and beamed away from the disk. This is possible if the new population is ejected from the base of a jet and may be part of a discrete jet. The X-ray flaring then could be due to synchrotron self-Compton (SSC) process as in the case of  radio-loud AGN. It is interesting to note the similarities of the differences between the X-ray spectra of radio-loud and radio-quiet AGN and that of I~Zw~1 in the low flux and high flux states. 

\subsection{The soft X-ray excess emission}
The high resolution soft X-ray spectrum of I~Zw~1, obtained with the RGS, is featureless and smooth. We did not detect any emission or absorption features below $1.5\kev$, the region of the spectrum usually affected by the warm absorbers. Thus, the RGS spectrum of I~Zw~1 supports the earlier observations with \chandra{} LETG and \xmm{} RGS that the soft X-ray excess emission of NLS1 galaxies is featureless and smooth (Turner et al. 2001; Collinge et al 2001; Gondoin et al. 2002; Page et al. 2003). 
Although the soft X-ray excess emission above a power-law continuum is well described by a blackbody component, its temperature is too high for the intrinsic emission from an accretion disk. The temperature of a standard thin accretion disk is 
\begin{equation} 
T(r) \sim 6.3\times 10^5 \dot{m}^{1/4} M_8^{-1/4} \left (\frac{r}{r_S} \right)^{-3/4} {\rm~ K}
\end{equation}
where $r_S$ is the Schwarzschild radius, and $M_8$ is the black hole mass in units of $10^8{\rm~M\odot}$ (Peterson 1997). The temperature at the last stable orbit ($r=3r_S$) for a Schwarzschild black hole is $\sim 35\ev$ for a black hole mass of $2.4\times 10^7{\rm~M\odot}$ accreting at the Eddington rate. This is an upper limit to the disk temperature, because $\dot{m} < 1$, and outer regions of the disk will have lower temperatures.

A more realistic model involving ionized reflection in addition to the blackbody and power-law components describes the $0.6-12\kev$ spectrum well. The reflection component fits the iron edge and contributes significantly to the soft excess emission thereby reducing the strength of the blackbody component. However, the blackbody temperature is still too high for an accretion disk.  

The primary X-ray continuum of Seyfert type AGN is thought to arise from thermal Comptonization of soft photons from the disk by the hot electrons in the corona (Haardt \& Maraschi 1991). Such a comptonized spectrum deviates from the power-law form at low energies, in addition to the high energy cut-off, due to the Maxwellian distribution of hot electrons (see e.g., Titarchuk 1994). Therefore, power-law is not an appropriate continuum to characterize the soft X-ray excess component. In the case of I~Zw~1, the soft X-ray excess emission above a `thermally comptonized continuum' is well described by ionized reflection and blackbody. Most importantly the temperature of the blackbody ($kT \sim 40\ev$) is consistent with that expected for the inner regions of an accretion disk around a $\sim 10^7{\rm~M\odot}$ black hole. 
Thus the soft X-ray emission of the proto-type NLS1 galaxy I~Zw~1 is not unusual in terms of strength, temperature or variability time scale.

\subsection{Iron features}
We detected an iron K$\alpha$ line at $\sim 6.7\kev$ and an iron K-edge at $\sim8.5\kev$. Both the features arise from He-like iron and suggest that the accretion disk of I~Zw~1 is ionized. Evidence for ionized accretion disks have been found in many NLS1 galaxies e.g., Ton~S180 (Vaughan et al. 2002; Comastri et al. 1998) Mrk~335 (Gondoin et al. 2002), Akn~564 (Vaughan et al. 1999) and are consistent with the general picture that accretion disks of AGN with steep X-ray spectrum are ionized (Dewangan 2002). The ionization stage of photo-ionized
material is determined by the ionization parameter,
$\xi=\frac{L}{nR^2}$, where $L$ is the luminosity of ionizing photons,
$n$ is
the density of the material, and $R$ is the distance of the
material from the ionizing source (Kallman \& McCray 1982). For
standard thin disks, Matt et al. (1993) showed that the ionization
parameter strongly depends on the accretion rate relative to the
Eddington rate, $\xi \propto ({\frac{L}{L_{Edd}}})^3$.  In order to
ionize He-like iron, $\xi$ should be at least $\sim500$ or
$\frac{L}{L_{Edd}}\gtrsim 0.2$ (Matt et al. 1993). This result is
consistent with the best-fit value of the ionization parameter derived from ionized reflection model.
These observations have important implications to the nature of the soft X-ray excess emission as reflection from ionized material contributes substantially in the soft X-ray band, as revealed in the case of I~Zw~1.  

\section{Conclusions}
We presented timing and spectral characteristics of the proto-type NLS1 galaxy I~Zw~1 based on an \xmm{} observation. The main results are as follows.
\begin{enumerate}
\item The $0.3-12\kev$ spectrum of I~Zw~1, obtained with \xmm{}, consists of four intrinsic spectral components, namely a steep ($\Gamma_X \sim 2.3$) primary continuum, a soft X-ray excess component described by a blackbody ($kT \sim 40\ev$) and ionized reflection from the disk, iron features -- K$\alpha$ line and K-edge both from He-like iron arising due to reflection of the primary continuum, and an intrinsic photoelectric absorption component.  
\item I~Zw~1 showed a large X-ray flare with changes in the luminosity $\Delta L \sim 6\times 10^{43}{\rm~erg~s^{-1}}$ on a time scale of $\sim 2800\s$. The radiative efficiency, $\eta = 0.01\pm0.002$, inferred from the flare is consistent with X-ray emission outside the last stable orbit around a Schwarzschild black hole.
\item X-ray spectrum of I~Zw~1 hardened with flux during the observation. 
This behavior  is uncommon among Seyfert galaxies. The hardening of the $0.6-10\kev$ 
is not only due to increase in the hard  ($2-10\kev$) band flux but also due to hardening 
of the power-law component at higher fluxes. The soft excess emission did not respond 
to the changes in the power-law continuum. The spectral variability of I~ZW~1 is difficult 
to understand in the framework of disk-corona models but appears to be related with ejection 
of relativistic plasma at the base of a jet. Simultaneous X-ray and radio observations 
would be required to understand such spectral variability during large X-ray flares.
\end{enumerate}

\acknowledgements This work is based on
observations obtained with \xmm{}, an ESA science mission with
instruments and contributions directly funded by ESA Member States and
the USA (NASA). REG acknowledges NASA award NAG5-9902 in support of
his Mission Scientist position on \xmm{}.

\clearpage

\begin{figure*}
\centering
\includegraphics[width=7cm,angle=-90]{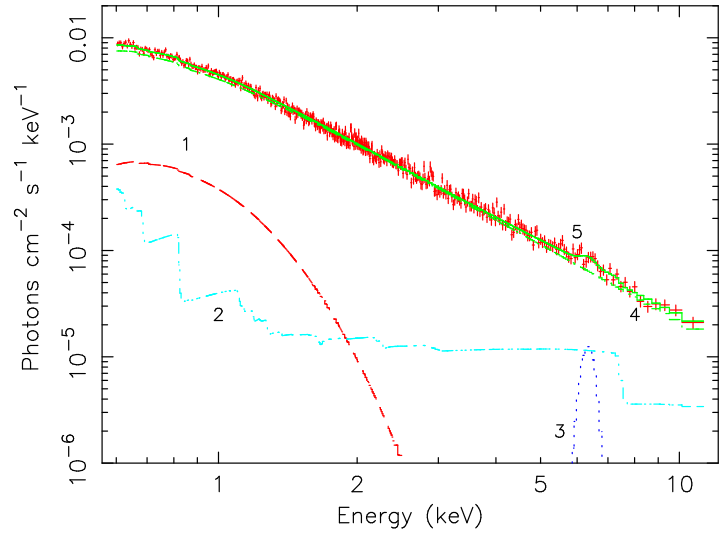}
\caption{Unfolded spectrum of I~Zw~1 and the best-fit model. For clarity, only PN data is shown. The model component are absorbed (1) blackbody ($kT \sim 200\ev$), (2) reflection from an ionized accretion disk, (3) a Gaussian line, and (4) a power-law.}
\label{f7}
\end{figure*}
                                                                                                      
\begin{figure*}
\centering
\includegraphics[width=7cm,angle=-90]{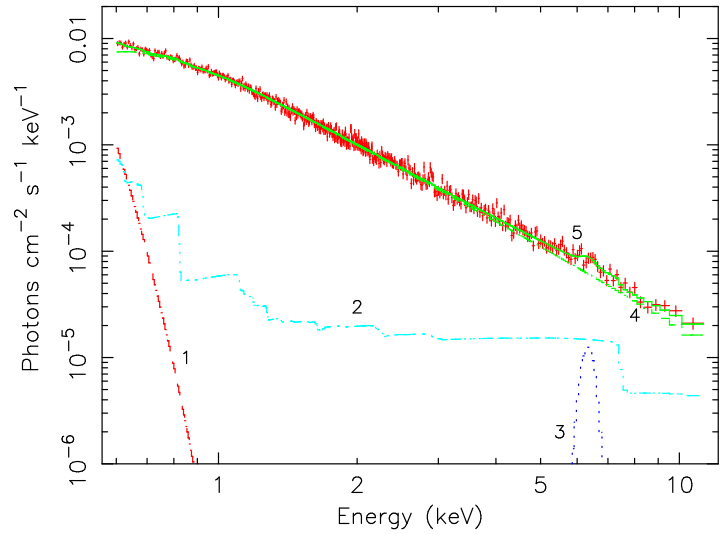}
\caption{Unfolded PN and MOS2 spectra and the best-fit model consisting of a (1) blackbody ($kT \sim 40\ev$), (2) reflection from an ionized accretion disk, (3) a Gaussian line, and (4) thermal Comptonization model ({\tt comptt}). }
\label{f8}
\end{figure*}
                                                                                                      
\begin{table*}
\caption{Best-fit spectral model parameters derived from the joint spectral fitting of the PN and MOS2 spectra of I~Zw~1.}
\label{tab1}
\begin{tabular}{lcccc}
\tableline\tableline
      &          & \multicolumn{3}{c}{Spectral model\tablenotemark{a}} \\
Comp. & Parameter    & BB+PL+G+Edge &  BB+PL+G+Refl & BB+G+CompTT+Refl \\
      &              & (model 1)   & (model 2) & (model 3) \\ 
\tableline
Excess Absorption & $\Delta \NH$ ($\times10^{20}{\rm~cm^{-2}}$ & $5.1_{-1.6}^{+1.7}$ & $8.5_{-2.2}^{+2.6}$  & $2.2_{-1.6}^{+3.5}$ \\
Blackbody & $kT$ ($\ev$)  &  $234.5_{-31.2}^{+28.7}$ & $199_{-51}^{+39}$ & $38_{-12}^{+45}$ \\
          & $n_{bb}\tablenotemark{b}$  & $2.3_{-0.5}^{0.5}$ & $1.8_{-0.9}^{+0.9}$ & $2278_{-2268}^{+2.5\times 10^5}$ \\ 
          & $f_{bb}(0.6-2\kev)\tablenotemark{f}$ & $0.11$  & $0.08$ & $0.007$  \\
Power law & $\Gamma$ &   $2.27_{-0.05}^{+0.05}$ & $2.41_{-0.09}^{+0.06}$ & -- \\
          & $n_{pl}\tablenotemark{c}$   &  $5.56_{-0.39}^{+0.35}$  & $6.3_{-0.4}^{+0.46}$ & -- \\
          & $f_{pl}(0.6-2\kev)\tablenotemark{f}$ & $0.92$ & $1.0$ & -- \\
          & $f_{pl}(2-10\kev)\tablenotemark{f}$ & $0.88$  & $0.85$ & -- \\   
CompTT    & $kT_{seed}\ev$ & -- & -- &$155_{-20}^{+5}$  \\
          & $kT_{plasma}\kev$ & -- & -- & $12.2_{-6.3}^{+44.8}$ \\
          & $\tau$  & -- & -- & $1.6_{-0.3}^{+3.1}$ \\
          & $n_{C}\tablenotemark{d}$ &  -- & -- & $1.3_{-0.7}^{+0.2}$  \\
          & $f_{C}(0.6-2\kev)\tablenotemark{f}$ & -- & -- &  $0.9$ \\
          & $f_{C}(2-10\kev)\tablenotemark{f}$  -- & -- &  $0.75$ \\
Gaussian  & $E_{line}\tablenotemark{e}$  & $6.72_{-0.20}^{+0.15}$ & $6.72_{-0.20}^{+0.22}$ & $6.72$ \\
          & $\sigma\tablenotemark{e}$ (eV) & $267_{-149}^{+307}$ & $225_{-225}^{+112}$ & $225$ \\
          & $f_{line}\tablenotemark{g}$   & $1.0_{-0.4}^{+0.7}$ & $0.7_{-0.5}^{+0.6}$ & $0.7$ \\
          & EW (eV) & $130$ & $89$ & $88$ \\
Edge      & $E_{edge}$  & $8.55_{-0.84}^{+0.35}$ & -- & --\\
          & $\tau$      & $0.14_{-0.11}^{+0.12}$ & -- & -- \\
Pexriv    & $R$         & -- & $0.65_{0.47}^{+0.67}$ & $0.83_{-0.12}^{+0.12}$ \\
          & $\xi$ ($\rm~erg~cm~s^{-1}$)     & -- & $1683_{-1624}^{+3066}$ & $1683$ \\
Total     & $f_{obs}(0.6-10\kev)\tablenotemark{f}$ & 1.6 & 1.6 & 1.6 \\
          & $f_{intr}0.6-10\kev)\tablenotemark{f}$ & 1.9 & 2.0 & 1.9 \\
          & $\chi^2_{min}/dof$ & 822/780 & 820.8/779 & 815.2/783 \\ 
\tableline
\end{tabular}
\tablenotetext{a}{BB - blackbody; PL - power law; G - Gaussian; Edge - absorption edge; Refl - ionized reflection model ({\tt pexriv}); CompTT - Thermal comptonization model. The spectral models were modified by the Galactic ($N_H^{Gal} = 5\times 10^{20}{\rm~cm^{-2}}$) and intrinsic absorption ($\Delta N_H$).}
\tablenotetext{b}{Blackbody normalization in units of $10^{-5}\times 10^{39}{\rm~erg~s^{-1}/(d/10~{\rm~kpc})^2}$, where $d$ is the distance.}
\tablenotetext{c}{Power-law normalization in units of $10^{-3}{\rm~photons~cm^{-2}~s^{-1}~keV^{-1}}$ at $1\kev$.}
\tablenotetext{d}{CompTT normalization in units of $10^{-3}{\rm~photons~cm^{-2}~s^{-1}~keV^{-1}}$ at $1\kev$.}
\tablenotetext{e}{Error on this parameter was calculated after fixing the edge parameters at their best-fit values.}
\tablenotetext{f}{In units of $10^{-11}{\rm~erg~cm^{-2}~s^{-1}}$}
\tablenotetext{g}{Iron line flux in units of $10^{-5}$}
\end{table*}

\begin{table*}
\caption{Results of time resolved spectral fitting of the PN and MOS2 spectra of I~Zw~1}
\label{tab2}
\begin{tabular}{lccccc}
\tableline\tableline
   &     & \multicolumn{4}{c}{BB+G+PL\tablenotemark{a}} \\
Comp. & Parameter    & Spectrum 1 &  Spectrum 2 & Spectrum 3 & Spectrum 4 \\
\tableline
Excess absorption & $\Delta \NH$ ($\times10^{20}{\rm~cm^{-2}}$ & $3.7_{-2.8}^{+2.9}$ & $6.2_{-3.5}^{+4.6}$ & $4.9_{-2.8}^{+2.8}$ & $3.4_{-3.3}^{+3.3}$   \\
Blackbody & $kT$ ($\ev$) & $244.5_{-41.7}^{+37.0}$ &  $201.1_{-44.1}^{+50.3}$ & $298.3_{-84.0}^{+53.3}$ & $275.8_{-63.2}^{+43.4}$ \\ 
          & $n_{bb}\tablenotemark{d}$  & $2.7_{-1.0}^{+0.9}$ & $3.1_{-2.2}^{+3.1}$ & $2.0_{-1.4}^{+1.2}$ & $2.1_{-0.8}^{+1.0}$  \\
          & $f_{bb}(0.6-2\kev)\tablenotemark{b}$ & $0.13$ & $0.13$ &  $0.11$ & $0.13$ \\ 
Power law & $\Gamma$ &   $2.28_{-0.08}^{+0.09}$ & $2.31_{-0.08}^{+0.08}$ & $2.25_{-0.09}^{+0.08}$ & $2.22_{-0.10}^{+0.09}$ \\
          & $n_{pl}\tablenotemark{c}$   &  $4.8_{-0.6}^{+0.6}$ & $5.2_{-0.6}^{+0.5}$ & $6.8_{-0.8}^{+0.9}$ & $4.6_{-0.8}^{+0.7}$\\
          & $f_{pl}(0.6-2\kev)\tablenotemark{g}$ & $0.79$ & $0.85$ & $1.12$ & $0.76$ \\
          & $f_{pl}(2-10\kev)\tablenotemark{g}$ & $0.71$ &  $0.74$ & $1.06$ &  $0.74$ \\
Gaussian  & $E_{line}\tablenotemark{d}$  & $6.71(f)$ & $6.71(f)$ & $6.71(f)$ & $6.71(f)$ \\
          & $\sigma\tablenotemark{d}$ (eV) & $266.5(f)$ & $266.5(f)$ & $266.5(f)$ & $266.5(f)$ \\
          & $f_{line}\tablenotemark{h}$   & $7.7_{-7.2}^{+8.6}$ & $15.6_{-10.1}^{+7.1}$ & $10.8_{-9.2}^{+9.8}$ &  $5.3_{-5.3}^{+9.6}$ \\
          & EW (eV)  &  $117$ &  $231$ & $107$ & $76$ \\
Edge      & $E_{edge}$  &  $8.64(f)$ &  $8.64(f)$ & $8.64(f)$ & $8.64(f)$ \\
          & $\tau$      & $0.19 (f)$  &  $0.19 (f)$  & $0.19 (f)$ & $0.19 (f)$  \\
Total     & $f_{obs}(0.6-10\kev)\tablenotemark{g}$ & $1.46$ & 1.50 & $2.04$ & $1.45$  \\
          & $f_{intr}0.6-10\kev)\tablenotemark{g}$ & $1.64$ & 1.73 & $2.31$ & $1.62$ \\
          & $\chi^2_{min}/dof$ & 588.2/612 & 556.2/600 &  726.6/731 & 709.3/648 \\
\tableline
\end{tabular}
\tablenotetext{a}{A combination of a blackbody (BB), power-law (PL), Gaussian line (G), and an absorption edge (Edge) modified by the Galactic absorption as well as intrinsic or excess absorption ($\Delta \NH$).} 
\tablenotetext{b}{Blackbody normalization in units of $10^{-5}\times 10^{39}{\rm~erg~s^{-1}/(d/10~{\rm~kpc})^2}$, where $d$ is the distance.}
\tablenotetext{c}{Power-law normalization in units of $10^{-3}{\rm~photons~cm^{-2}~s^{-1}~keV^{-1}}$.}
\tablenotetext{d}{The line energy and the Gaussian $\sigma$ were fixed to their best-fit values derived from the mean spectrum of I~Zw~1.}
\tablenotetext{f}{Fixed parameters}
\tablenotetext{g}{In units of $10^{-11}{\rm~erg~cm^{-2}~s^{-1}}$}
\tablenotetext{h}{Iron line flux in units of $10^{-6}$}
\end{table*}

\end{document}